\documentclass[twocolumn]{aastex631}
\usepackage{amsmath}

\begin{document}

\title{Real-time experimental demonstrations of a photonic lantern wavefront sensor}

\author[0000-0001-8542-3317]{Jonathan W. Lin}
\affiliation{University of California, Los Angeles, Physics \& Astronomy Department, Los Angeles, 475 Portola Plaza Los Angeles, CA 90095, USA}

\author[0000-0002-0176-8973]{Michael P. Fitzgerald}
\affiliation{University of California, Los Angeles, Physics \& Astronomy Department, Los Angeles, 475 Portola Plaza Los Angeles, CA 90095, USA}

\author[0000-0002-6171-9081]{Yinzi Xin}
\affiliation{California Institute of Technology, Department of Astronomy, Pasadena, CA 91125, USA}

\author[0000-0003-1392-0845]{Yoo Jung Kim}
\affiliation{University of California, Los Angeles, Physics \& Astronomy Department, Los Angeles, 475 Portola Plaza Los Angeles, CA 90095, USA}

\author{Olivier Guyon}
\affiliation{National Astronomical Observatory of Japan, Subaru Telescope, 650 North Aohoku Place, Hilo, HI 96720, USA}
\affiliation{The University of Arizona, Department of Astronomy and Steward Observatory, 933 N. Cherry Ave., Tucson, AZ 85719, USA}

\author{Barnaby Norris}
\affiliation{The University of Sydney, Sydney Institute for Astronomy, Physics Road, Sydney, NSW 2006, Australia}

\author{Christopher Betters}
\affiliation{The University of Sydney, Sydney Astrophotonic Instrumentation Laboratory, Sydney, NSW 2006, Australia}

\author[0000-0002-5606-3874]{Sergio Leon-Saval}
\affiliation{The University of Sydney, Sydney Astrophotonic Instrumentation Laboratory, Sydney, NSW 2006, Australia}

\author[0000-0002-1094-852X]{Kyohoon Ahn}
\affiliation{National Astronomical Observatory of Japan, Subaru Telescope, 650 North Aohoku Place, Hilo, HI 96720, USA}

\author[0000-0003-4514-7906 ]{Vincent Deo}
\affiliation{National Astronomical Observatory of Japan, Subaru Telescope, 650 North Aohoku Place, Hilo, HI 96720, USA}

\author{Julien Lozi}
\affiliation{National Astronomical Observatory of Japan, Subaru Telescope, 650 North Aohoku Place, Hilo, HI 96720, USA}

\author{Sébastien Vievard}
\affiliation{National Astronomical Observatory of Japan, Subaru Telescope, 650 North Aohoku Place, Hilo, HI 96720, USA}

\author{Daniel Levinstein}
\affiliation{University of California, Irvine, Department of Physics \& Astronomy, 4129 Frederick Reines Hall, Irvine, CA 92697, USA}

\author[0000-0001-6871-6775]{Steph Sallum}
\affiliation{University of California, Irvine, Department of Physics \& Astronomy, 4129 Frederick Reines Hall, Irvine, CA 92697, USA}

\author[0000-0001-5213-6207]{Nemanja Jovanovic}
\affiliation{California Institute of Technology, Department of Astronomy, Pasadena, CA 91125, USA}




\begin{abstract}
The direct imaging of an Earth-like exoplanet will require sub-nanometric wavefront control across large light-collecting apertures, to reject host starlight and detect the faint planetary signal. Current adaptive optics (AO) systems, which use wavefront sensors that reimage the telescope pupil, face two challenges that prevent this level of control: non-common-path aberrations (NCPAs), caused by differences between the sensing and science arms of the instrument; and petaling modes: discontinuous phase aberrations caused by pupil fragmentation, especially relevant for the upcoming 30-m class telescopes. Such aberrations drastically impact the capabilities of high-contrast instruments. To address these issues, we can add a second-stage wavefront sensor to the science focal plane. One promising architecture uses the photonic lantern (PL): a waveguide that efficiently couples aberrated light into single-mode fibers (SMFs). In turn, SMF-confined light can be stably injected into high-resolution spectrographs, enabling direct exoplanet characterization and precision radial velocity measurements; simultaneously, the PL can be used for focal-plane wavefront sensing. We present a real-time experimental demonstration of the PL wavefront sensor on the Subaru/SCExAO testbed. Our system is stable out to around $\pm$ 400\,nm of low-order Zernike wavefront error, and can correct petaling modes. When injecting $\sim$ 30 nm RMS of low order time-varying error, we achieve $\sim$ 10$\times$ rejection at 1 s timescales; further refinements to the control law and lantern fabrication process should make sub-nanometric wavefront control possible. In the future, novel sensors like the PLWFS may prove to be critical in resolving the wavefront control challenges posed by exoplanet direct imaging.
\end{abstract}



\section{Introduction} \label{sec:intro}

From the ground, planar wavefronts from distant stars are warped due to Earth's turbulent atmosphere, and can become further distorted by the imperfect and unstable optics of astronomical instruments. In astronomy, these challenges motivate adaptive optics (AO), a technique which combines wavefront sensors (WFSs) and deformable mirrors (DMs) to actively flatten incoming wavefronts. Using AO, modern observing facilities have imaged exoplanets, uncovered circumstellar environments, examined the galactic center, and probed the structure of active galactic nuclei. Beyond astronomy, AO finds applications in free-space optical communications, microscopy, and remote sensing, all of which must contend with the propagation of light through inhomogeneous and dynamic media.
\\\\
One of the primary goals for astronomy in the upcoming decade will be the direct imaging of an Earth-like exoplanet, and the characterization of any potential biological signatures: a scientific and technical feat whose importance was reiterated by the Astro 2020 decadal survey \citep{decadal}, and whose challenges are now driving the development of new technologies.
One of the primary hurdles is the suppression of light from the host star, which will outshine an Earth-like companion by a factor of $10^{10}$ in the visible \citep{Traub:10}. To do so, we require coronagraphs or external starshades to blot out starlight, paired with wavefront control which must flatten incoming wavefronts at sub-nanometer precision. Such stringent demands cannot be met by conventional AO, which use sensors (e.g. Shack-Hartmann, pyramid) located in a conjugate pupil separate from the science focal plane and therefore suffer from non-common-path aberrations (NCPAs): instrumental aberrations that appear exclusively either in the science or sensing arms of the instrument \citep{Martinez:12:NCPA1}. Additional complications include the low-wind effect (LWE), where the temperature gradients in a segmented aperture lead to aberrations that are discontinuous at segment boundaries \citep{NDiaye:18}, and petaling, in which the same kind of aberrations arise due to drifting misalignments in telescopes with fragmented primaries. Both are challenging to correct with general purpose sensors which re-image the pupil; petaling is of particular concern for the upcoming 30~m-class telescopes, which will provide the best chance of directly imaging Earth-like exoplanets from the ground.
\\\\
We can avoid these difficulties by using wavefront sensors that operate in the science focal plane. This can be enabled in two ways: through phase diversity (e.g. COFFEE \citep{COFFEE}, and related algorithms such as F\&F \citep{FF} and DrWho \citep{DRWHO}); or through purpose-built sensing optics such as phase holograms, asymmetric pupil masks, and dephased pinhole masks (e.g. cMWFS, APF-WFS, ZELDA, vAPP \citep{cMWFS,APFWFS,ZELDA,vAPP}), sometimes coupled with non-linear phase retrieval algorithms. In this work we consider an alternate architecture that works with standard linear phase retrieval algorithms, using a photonic lantern (PL): a slowly transitioning waveguide which efficiently couples multi-modal light into multiple single-moded outputs \citep{Leon-Saval:05,Birks:15}. Importantly, when used to couple the aberrated telescope beam in the focal plane, PLs already have utility in several high-contrast imaging applications, including spectro-imaging and starlight nulling. PLs are also uniquely suited to couple aberrated telescope light into highly stable diffraction-limited spectrometers \citep{Lin:21}, enabling direct exoplanet spectroscopy, and can serve as a gateway into a wider ecosystem of astrophotonic devices such as arrayed waveguide gratings and photonic integrated circuits \citep{roadmap}.
Simultaneously, low-spatial-frequency aberrations such as NCPAs and LWE modes can be sensed in the true science focal plane by monitoring the fluxes of the lantern's outputs. By construction, the photonic lantern wavefront sensor (PLWFS), at least in a single monochromatic channel, is a low-order sensor, with a maximum number of sensed modes equal to the number of lantern ports (or twice that if polarizations can be separated) \citep{Lin:22,Lin:23}. Accordingly, we envision that in practical ground-based applications the PLWFS will be most useful as a second-stage system, correcting the low-order NCPAs and petalling aberrations left over by a first-stage pupil-plane WFS control loop. 
\\\\
Previous numerical modelling \citep{Lin:22,Lin:23} and experimental results \citep{Corrigan:18,Norris:20} have made important progress in developing the PLWFS. However, unknowns, such as the PLWFS's linearity, dynamic range, and stability, especially as part of a real-time AO system, have so far prevented the PLWFS from being realistically considered. The next step --- a real-time demonstration of the PLWFS as part of a modern AO system --- would verify the PLWFS as a future pathway to wavefront sensing in the true science focal plane, one that is further unique due to the non-WFS capabilities that PLs can simultaneously provide, including starlight nulling, spectro-imaging, and high-resolution spectroscope injection. In this work, we take this step, using the SCExAO testbed at Subaru telescope. Our results place the PLWFS firmly on the path to eventual integration with the next generation of astronomical instruments, which will ultimately enable the imaging and characterization of an Earth-like exoplanet.

\begin{figure*}[ht]
    \centering
    \includegraphics[width=\textwidth]{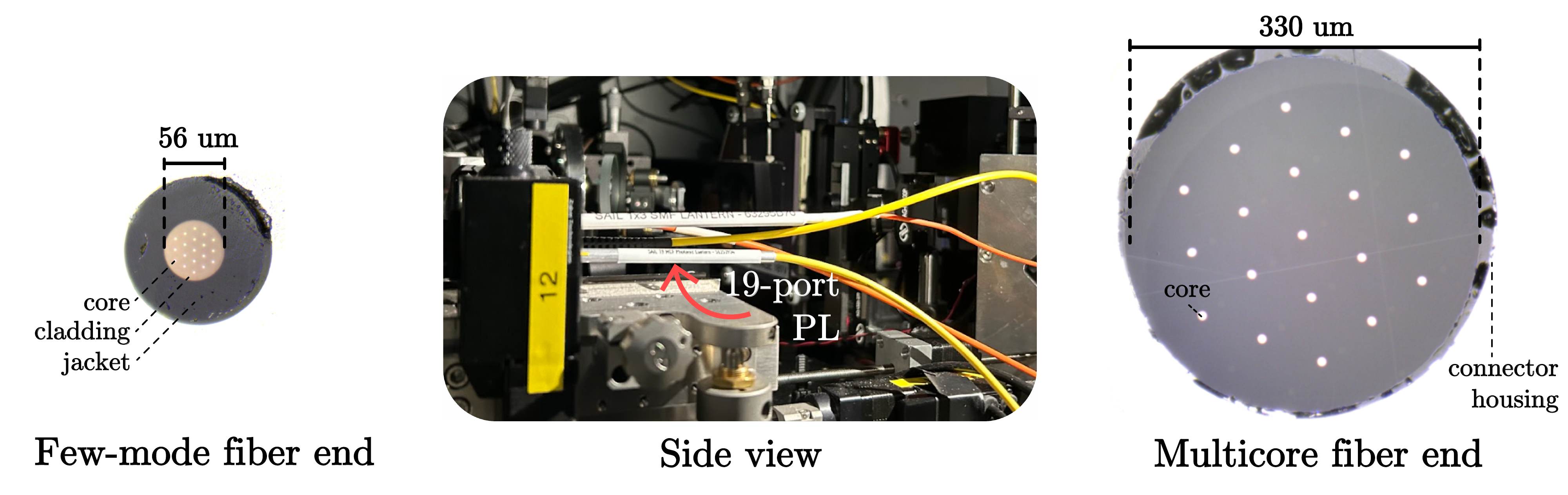}
    \caption{End face and side views of the 19-port PL used for our wavefront sensing tests. End views are taken using a microscope at 792$\times$ magnification, and are approximately to scale relative to each other. To increase the visibility of the single-mode cores, visible light was injected through the lantern during imaging for the multicore end, and reverse injected during imaging of the few-mode fiber end.}
    \label{fig:0}
\end{figure*}
\section{Methodology}
\subsection{Testbed setup}
We used the near-infrared testbed on SCExAO \citep{Lozi:20,Jovanovic:16} to demonstrate real-time control with the PLWFS. Microscope images of the 19-port PL used in this work, as well as a picture of the lantern mounted in the SCExAO testbed, are shown in Figure \ref{fig:0}. A simplified diagram of the SCExAO beampath, containing only optics relevant for our tests, is given in the top panel of Figure \ref{fig:1}, while an overview of the closed loop calibration and control process is shown at the bottom. Our light source is a supercontinuum white light laser, with a narrowband 1550 $\pm$ 50 nm filter, which is then collimated onto a $50\times 50$ actuator DM and apodized using a set of Gaussian beam-shaping lenses. We use these lenses only because the downstream optics of the fiber injection unit were sized for the smaller apodized beam. The beam is then divided by a 90:10 beam splitter, which sends 10\% of the light through to SCExAO's internal NIR camera (FLI C-Red 2) for monitoring of the point-spread function. The remaining 90\% of the light is sent into the fiber injection unit: a 4-axis translation stage (shown in Figure \ref{fig:0}, middle) that allows us to change the focal ratio of the injection as well as reposition the lantern in the focal plane. Finally, the 19 single-mode lantern output spots are imaged onto a detector (FLI C-Red 1).

\subsection{Software}
\begin{figure*}
    \centering
    \includegraphics[width=\textwidth]{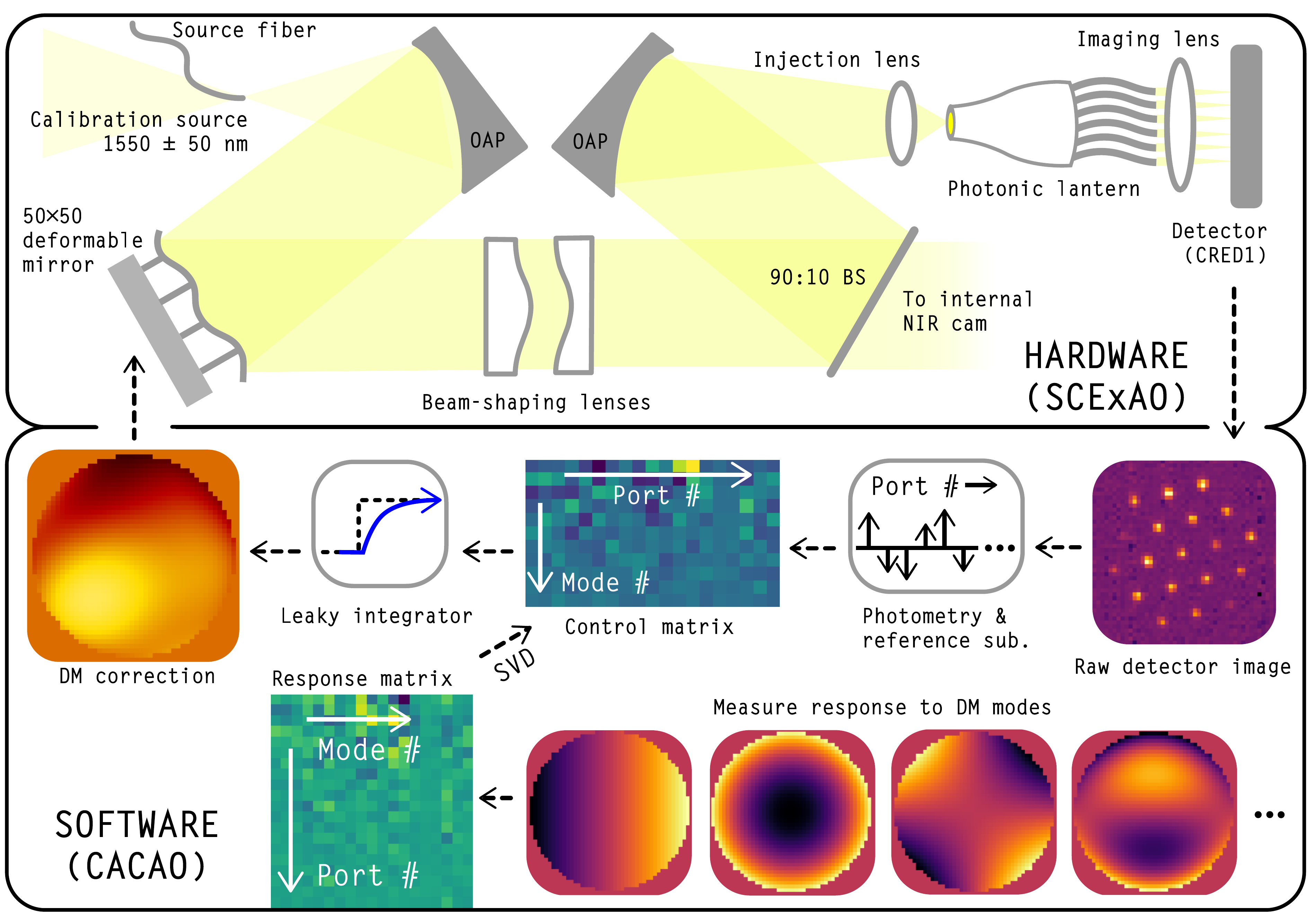}
    \caption{Top: basic hardware setup for PL wavefront sensing tests on SCExAO. OAP and BS stand for off-axis parabola and beam splitter, respectively. Only the relevant components of the SCExAO test bench are shown. Bottom: overview of closed loop control for the PLWFS, including intermediate steps like response matrix measurement. Loop calibration and control is handled with the CACAO package.}
    \label{fig:1}
\end{figure*}

To calibrate and close the PLWFS loop, we use the Compute And Control for Adaptive Optics (CACAO) package. Figure \ref{fig:1} (bottom) gives an overview of the software steps required to close the loop. The raw frame taken from the CRED1 detector must be processed before it can be used for AO control. This includes dark subtraction, photometry extraction, image normalization, and reference subtraction. Initial data processing converts the full-frame detector image into a 19-dimensional vector, which we then multiply against the control matrix. The output is then fed into a leaky integrator; the output mode values are converted back to a $50\times 50$ displacement map which is then sent out to the SCExAO DM. To compute the control matrix, we measure the response of the PLWFS to a set of aberration modes (Zernike, petaling, Hadamard, etc.) and calculate the pseudo-inverse, setting our regularization parameter to 0.1. The loop was run at 1 kHz, but in principle could be run even faster; because the computations for linear phase retrieval with only a few modes are lightweight, control speed is limited by hardware, not software.

\subsection{Photonic lantern}
The lantern in our demonstration was manufactured at the Sydney Astrophotonics Instrumentation Laboratory (SAIL) using a tapered-fiber process: a custom multicore fiber with 19 single-mode channels (hexagonal array, 6.5 $\mu$m core diameter, 60 $\mu$m core spacing, numerical aperture 0.14) was inserted into a lower-index fluorine glass capillary, and then one end was heated and drawn to form a tapered structure. The left and right panels of figure \ref{fig:0} show the endface geometries of the PL. 

\section{Results}\label{sec2}

\subsection{Closed-loop correction of static aberrations}
\begin{figure*}
    \centering
    \includegraphics[width=\textwidth]{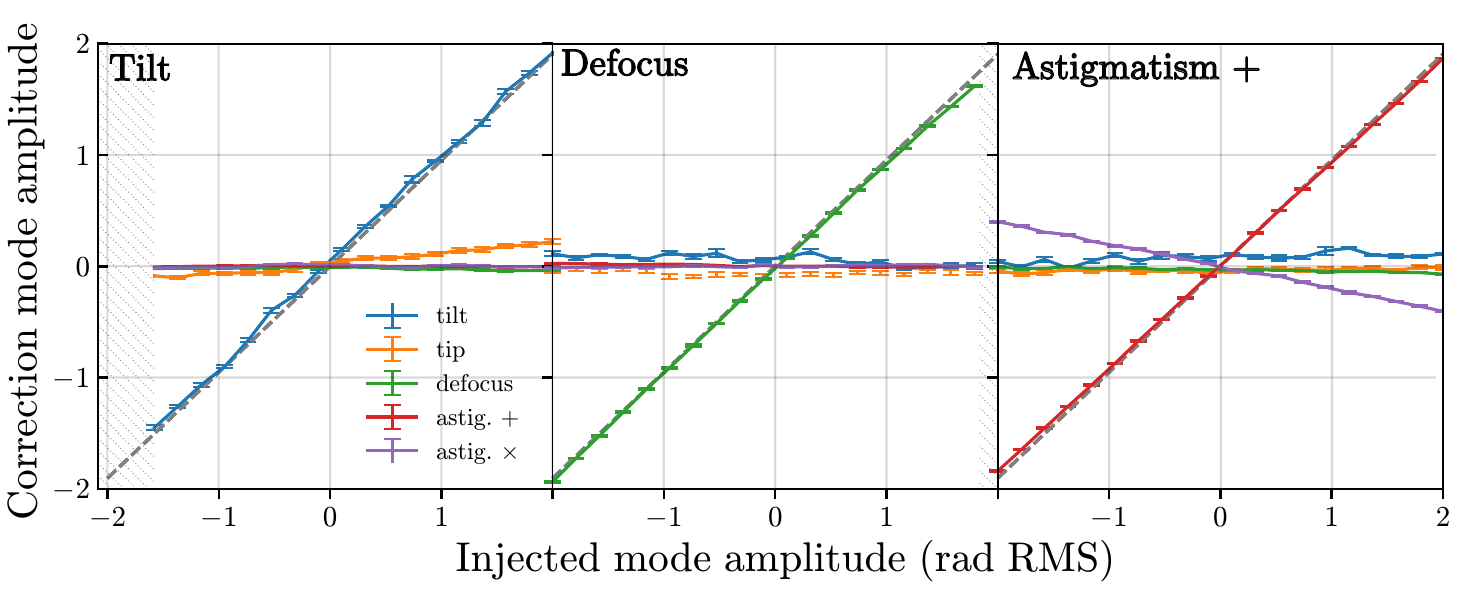}
    \caption{Correction applied by closed loop, as a function of the amount of static Zernike aberration injected into the system. The sign of the $y$ axis has been flipped for clarity (in a perfect system, the injection and correction mode amplitude would sum to 0). The left, middle, and right panels show the loop behavior when injecting tilt, defocus, and astigmatism, respectively. For each injected mode amplitude, 20 measurements of the closed loop correction were taken; vertical bars show the standard deviation. The diagonal dashed grey lines show the line $y=0.95x$, following the expected correction of a static aberration for our chosen loop parameters. Hatching shows regions where the loop becomes unstable. }
    \label{fig:2}
\end{figure*}

\begin{figure*}
    \centering
    \includegraphics[width=\textwidth]{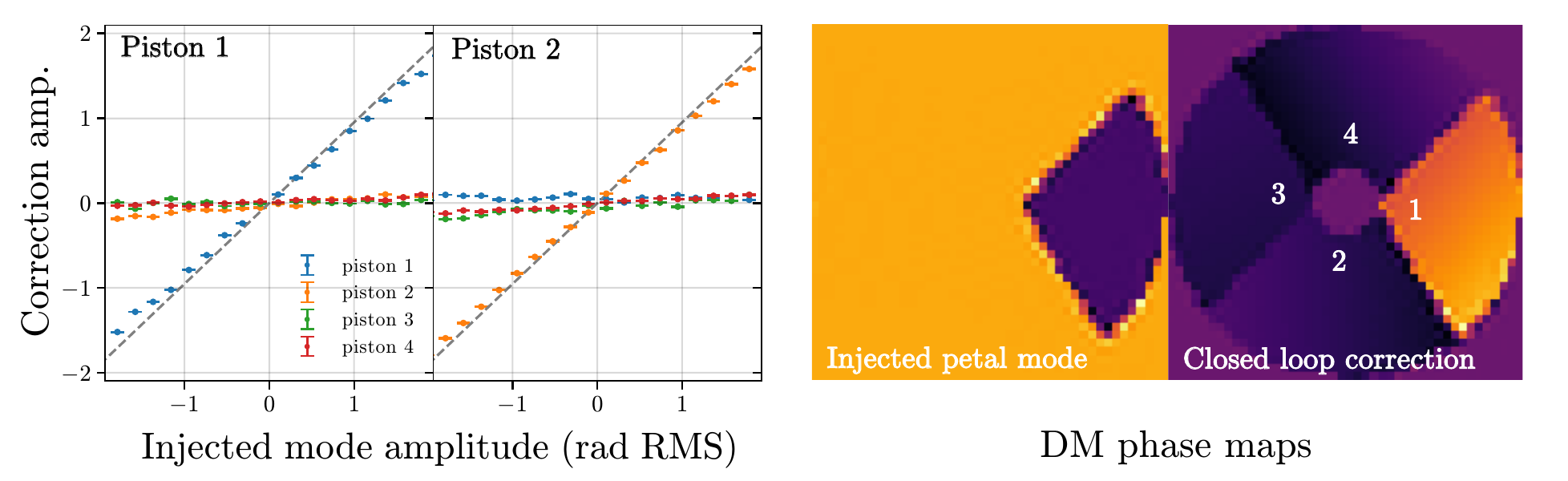}
    \caption{Left: correction applied by closed loop, as a function the amount of static LWE piston aberration artificially injected into the system. We only show correction of the first two piston modes for brevity; the correction of the other two modes behaves similarly. Right: a comparison between an example LWE piston mode and the associated correction applied by the PLWFS loop. Numbers in the rightmost panel show how the piston segments are indexed.  }
    \label{fig:2.5}
\end{figure*}

We first closed the loop on the first 5 non-piston Zernike modes, at a frequency of 1 kHz using a leaky integrator control law, with ${\rm leak} = 0.99$ and ${\rm gain} = 0.2$. We choose this few-mode, low-spatial-frequency basis both because the PLWFS is a few-mode sensor, and because most of the power in instrumental aberrations will appear in the first few Zernike modes \citep{Sauvage:07}. To test the loop, we injected a fixed amount of a single Zernike mode, scanning in both mode amplitude and mode index. At each point in the scan, after injecting the WFE and letting the loop settle, we sampled the closed-loop correction 20 times over the course of 2 seconds. This procedure is idealized in the sense that it neglects the effect of AO residuals, which will degrade the performance of the sensor in real-world operation; nevertheless, we believe that a comprehensive treatment of WFS performance in the presence of realistic AO residuals (e.g. as was done in \cite{Engler:22} for the pyramid) outside the scope of this work. We plan on pursuing such characterizations with future on-sky tests.
\\\\
Figure \ref{fig:2} shows the correction of three Zernike modes over a range of static injected mode amplitudes; the other two modes show similar behavior and are omitted for brevity.
We note that the loop was able to consistently correct $\sim$95\% of the injected WFE, which is the expected amount of correction when setting ${\rm leak} = 0.99$ and ${\rm gain} = 0.2$ (see \S\ref{ap:b}). Furthermore, the loop remained stable out to around $\pm$1.6 radians of RMS WFE, or about 400 nm at our injection wavelength. The dynamic range was limited by tilt. However, it is important to note that precise loop properties are sensitive to lantern alignment, and that we did not apply a rigorous optimization to align the lantern in the best location for WFSing (which may not be at the lantern's center, or even in focus). Nevertheless, our estimate of dynamic range falls in the middle of the predicted lower and upper limits from prior simulations \citep{Lin:23}, which suggested that a 19-port lantern sensing the first 5 non-piston Zernikes would have good linearity out to at least 0.5 radians and would be limited by degeneracies, arising from nonlinearity, beyond 2.3 radians. Figure \ref{fig:2} also shows aliasing between the two astigmatism modes (and to a lesser extent, the two tilt modes), which was not shown in simulations; it is unclear if this result is specific to our particular 19-port lantern, or if it can be reduced by adjusting the alignment. 
\\\\
We also consider the correction of petaling/LWE modes, an additional source of aberration besides NCPAs which will be particularly problematic for the upcoming 30-m telescopes, and are difficult to sense with conventional pupil-plane sensors. The SCExAO pupil is divided by spiders into four aperture segments, giving a total of 12 LWE modes: a local piston and two local tilt modes per aperture segment. These 12 LWE modes correspond to 11 degrees of freedom, since overall piston has no effect. We find that out of these 11 degrees, our 19-port PLWFS is able to sense 8 of them, including all 4 segment pistons; more modes might be recovered by tweaking alignment or by using a higher mode-count lantern. Figure \ref{fig:2.5} (left) repeats the laboratory procedure of Figure \ref{fig:2} for two of the four LWE segment piston modes, while \ref{fig:2.5} (right) compares phase maps from the WFE injection and closed-loop correction channels of the DM. 
We find that the PLWFS is capable of tracking all four piston modes, but with a slight undercorrection of $\sim$5--10\%, perhaps due to cross-talk. Unlike our previous test, we observe no loop instabilities over the entire mode amplitude range of -2 to 2 radians, for any of the segment piston modes. 
\subsection{Dynamic aberrations}
\begin{figure}
    \centering
    \includegraphics[width=\columnwidth]{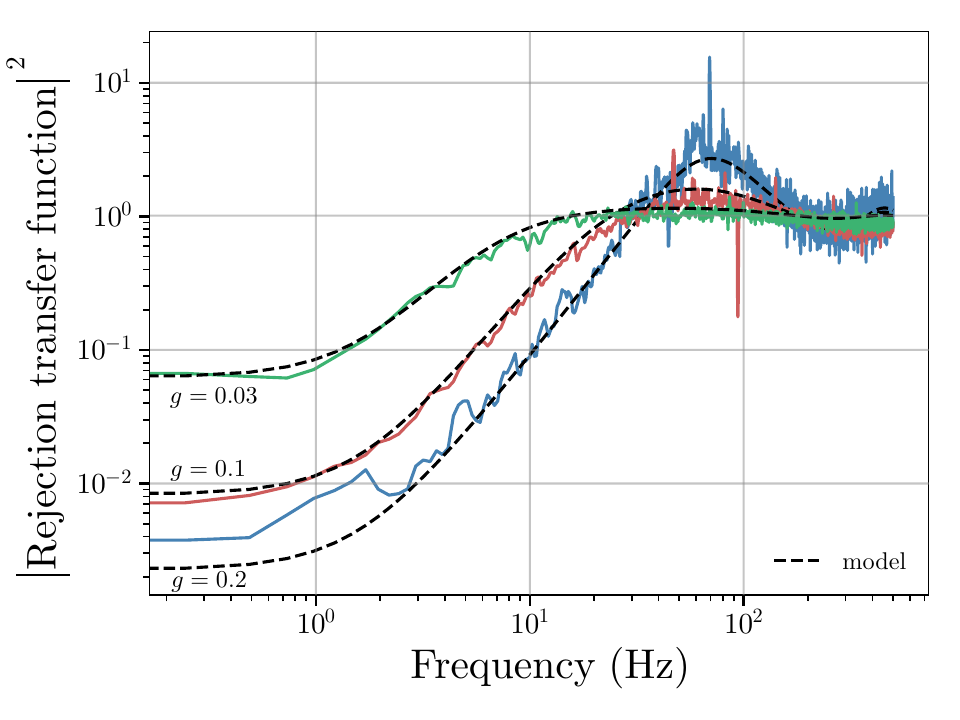}
    \caption{Colored lines: experimentally measured squared modulus of the AO system's rejection transfer function (equal to the ratio between the closed and open loop power spectral distributions) for the first control mode, at different leaky integrator gains. Transfer functions for other control modes look similar. Dashed black lines: theoretically expected squared modulus of the rejection transfer function, at different gains. }
    \label{fig:3}
\end{figure}

While the PLWFS can suppress static aberrations in closed-loop operation, in practice instrumental aberrations evolve temporally due to mechanical drifts, thermal expansion, and gravity vector changes caused by the slewing telescope. Excluding vibrations, the temporal evolution is typically on the timescale of seconds to minutes \citep{Martinez:12:NCPA1,Martinez:13:NCPA2}. Accordingly, we tested the loop next by injecting time-evolving low order WFE with a decorrelation timescale of 1\,s. This artificial WFE was composed from the first 7 non-piston Zernike modes (tilt, defocus, astigmatism, and coma); each mode amplitude is independently updated at a rate of 1 kHz according to an autoregressive formula, setting a per-mode amplitude of 0.05 radians RMS. For more details, see \S\ref{ap:a}. To calibrate the loop, we measured the sensor response matrix against the first 20 non-piston Zernike modes, and used a singular value decomposition to compute the control modes of the system. In all, we found 12 control modes, constructed from independent linear combinations of the first 20 Zernikes. Note that we cannot recover a full 19 modes with a 19-port PLWFS because we lose one degree of freedom to image normalization and another to a global piston-like mode \citep{Lin:22}; beyond that, there is no guarantee \textit{a priori} that the 19 lantern outputs will behave independently for a given aberration basis. We then logged the PLWFS output in both open and closed loop; the closed-loop leak was 0.99. The ratio of the open- and closed-loop power spectral densities (PSDs), which we estimated from the collected time-series data, approximates the squared modulus of the system's rejection transfer function. We plot the experimentally measured transfer functions for the first control mode in Figure \ref{fig:3} at different gains, along with the transfer functions expected by a theoretical model for our closed-loop system, set entirely by system latency, detector framerate, leak, and gain. Our model transfer function is presented in \S\ref{ap:b}.
All measured transfer functions in Figure \ref{fig:3} agree with models, and show successful rejection of the slowly varying components of the injected WFE. We find 0 dB frequencies of 15, 23, and 32 Hz, and rejection at 1 Hz of roughly $4\times$, $10\times$, and $12\times$, for gains of 0.03, 0.1, and 0.2, respectively. We also note that these transfer functions can be recovered by comparing open- and closed-loop PSDs, even without injecting artificial WFE (relying solely on the actual system instabilities), implying that the PLWFS can actively correct the ambient WFE of the SCExAO testbed. 

\subsection{Stability}
In a perfectly stable AO system, measurement of the response matrix only needs to happen once; in reality, an accumulation of slight changes in WFS optical properties (e.g. caused by temperature drifts) will require eventual re-calibration. In this subsection, we present a rudimentary estimate of the average drift rate of the PL response matrix, separate from drift inherent to the SCExAO testbed. The basis of our test is as follows. Suppose at some initial time, the response matrix of the PLWFS, $A_0$, is measured, and the control matrix $A_0^+$ is computed. Then, $A_0^+ A_0 = I$, where $I$ is the identity matrix. However, later measurements of the response matrix, denoted $A_n$, will deviate from $A_0$, and thus the product $A_0^+A_n$ will drift from $I$. This accumulated miscalibration error, err$_n$, can be quantified as 
\begin{equation}
    {\rm err}_n = |A_0^+A_n-I|    
\end{equation}
where $|M|$ denotes the Frobenius norm of matrix $M$.
\\\\
We applied this test to the PLWFS, over the course of one week. In practice, we first measured the reference response and control matrices, $A_0$ and $A_0^+$, for 11 control modes. At roughly the same time on later days, we then re-aligned the PL injection, closed the loop using the reference control matrix, and used \texttt{cacao}'s built-in \texttt{selfRM} function to measure $A_0^+A_n$ directly. \texttt{selfRM} works by manually injecting a small amount of each control mode, in sequence, into the system while the loop is closed, and recording the applied correction. After one week, we compute an accumulated miscalibration error of $\sim 1.4$. On average, this corresponds to an accumulated per-mode error of 1.4/11 $\approx$ 0.13 (13\%, since each column of $A_0^+A_n$ should have a norm of 1) over the course of the week, or a per-mode drift of around 2\% per day. This measurement of long-term stability (separate from diurnal stability) is a lower bound since we currently cannot completely compensate for the drifting bulk optics on the SCExAO testbed. Thus, the accumulated miscalibration error we measure is at least partially attributable to testbed instability; nevertheless, our results confirm that we can still close the wavefront control loop using a week-old calibration. 

\section{Discussion and Conclusion}
The PLWFS offers a novel approach to focal-plane wavefront sensing, enabling both the correction of NCPAs and petaling modes, which currently limit achievable contrast in exoplanet direct imaging, as well as a host of additional high-contrast applications such as starlight nulling and high-resolution spectrometer injection. We present the first real-time demonstration of the photonic lantern wavefront sensor, and verify its ability to correct both low-order Zernike modes and LWE/petaling modes from the science path focal plane. We further confirm the system's ability to track and correct dynamic Zernike WFE which varies on a timescale of $\sim 1$s, representative of quasistatic NCPAs --- currently one of the main limiting factors in achieving higher contrasts. Our control method has the added benefit of simplicity, using conventional linear phase retrieval and thus easily integrable into existing AO systems. However, future work still needs to be done before a PLWFS can be adopted in next-generation high-contrast imaging instruments. For one, we require a better understanding of how PL geometry and the overall system alignment impacts the properties of the AO system. In terms of practicality, we must also measure the PLWFS's WFE sensitivity and limiting magnitude in realistic conditions. In regards to the latter, we expect that the PLWFS should be able to work with fainter guide stars than conventional pupil-plane AO systems, because PLs have relatively high coupling efficiencies and concentrate collected light into fewer pixels. In the near future, we hope to take the PLWFS on-sky, and to leverage spectral dispersion and/or polarization to increase the amount of wavefront information provided by the lantern. Eventually, we believe that the PLWFS, and the wider ecosystem of astrophotonic devices, will ultimately play a critical part in the successful direct imaging and characterization of another Earth.
 
\begin{acknowledgments}
This material is based upon work supported by the National Science Foundation Graduate Research Fellowship Program under Grant No. DGE-2034835. Any opinions, findings, and conclusions or recommendations expressed in this material are those of the author(s) and do not necessarily reflect the views of the National
Science Foundation. This work was also supported by the National Science Foundation under Grant Nos. 2109232 and 2308361.
\end{acknowledgments}

%



\appendix
\section{Simulating temporal WFE}\label{ap:a}
To simulate dynamic WFE, we continuously apply a dynamic phase map composed from the first 7 non-piston Zernike modes (tilt, defocus, astigmatism, and coma) to a DM channel separate from the closed loop channel. Denote $z^j_n$ as the amplitude of mode $j$ at timestep $n$. We use the following autoregressive formula to update the mode amplitudes:
\begin{equation}
    z_j^{n+1} = e^{-1/t_c}\, z_j^n + a_j\,\sqrt{1-e^{-2/t_c}}\,m^n.
\end{equation}
Here, $n$ superscripts denote timestep while $j$ subscripts denote mode index. $t_c$ is the decorrelation time in frames, $a_j$ is the desired average mode amplitude for mode $j$ in units of radians RMS, and $m^n$ is the $n$th draw of a Gaussian distributed random variable with mean $\mu=0$ and standard deviation $\sigma=1$. In this particular test, we set $t_c = 1000$ frames and $a_j = 0.05$ radians RMS, for all modes. The overall phase map was updated at a rate of 1\,kHz.

\section{Modelled rejection transfer function}\label{ap:b}
A classic AO loop using linear phase retrieval and a leaky integrator control law has the following complex-valued rejection transfer function for mode $j$:
\begin{equation}
    h_{{\rm rej},j}(f) = \left[1 + g_j\dfrac{{\rm sinc }(T_e f)e^{-\pi i f(T_e+2\tau)}}{ 1 - l_j\, e^{-2\pi i T_e f}}\right]^{-1}.
\end{equation}
Here, ${\rm sinc}(x)\equiv \sin(\pi x)/\pi x$. This transfer function model is determined by four parameters: the modal gain $g_j$, the modal leak $l_j$, the detector integration time $T_e$, and the loop latency $\tau$. The leakage term $l_j$ is defined such that $l_j = 1$ corresponds an ordinary (i.e. non-leaky) integrator. For our PLWFS tests on SCExAO, the detector integration time is 1\,ms and the loop latency was independently measured using CACAO to be 2.7\,ms. The leak was set to 0.99 for all tests.
\\\\
For static aberrations, the rejection becomes
\begin{equation}
 h_{{\rm rej},j}(f=0) = \dfrac{1-l_j}{1-l_j+g_j}.
\end{equation}
\bibliography{sample631}{}

\begin{thebibliography}{}
\expandafter\ifx\csname natexlab\endcsname\relax\def\natexlab#1{#1}\fi
\providecommand{\url}[1]{\href{#1}{#1}}
\providecommand{\dodoi}[1]{doi:~\href{http://doi.org/#1}{\nolinkurl{#1}}}
\providecommand{\doeprint}[1]{\href{http://ascl.net/#1}{\nolinkurl{http://ascl.net/#1}}}
\providecommand{\doarXiv}[1]{\href{https://arxiv.org/abs/#1}{\nolinkurl{https://arxiv.org/abs/#1}}}

\bibitem[{Birks {et~al.}(2015)Birks, Gris-S\'{a}nchez, Yerolatsitis, Leon-Saval, \& Thomson}]{Birks:15}
Birks, T.~A., Gris-S\'{a}nchez, I., Yerolatsitis, S., Leon-Saval, S.~G., \& Thomson, R.~R. 2015, Adv. Opt. Photon., 7, 107, \dodoi{10.1364/AOP.7.000107}

\bibitem[{{Bos, S. P.} {et~al.}(2019){Bos, S. P.}, {Doelman, D. S.}, {Lozi, J.}, {Guyon, O.}, {Keller, C. U.}, {Miller, K. L.}, {Jovanovic, N.}, {Martinache, F.}, \& {Snik, F.}}]{vAPP}
{Bos, S. P.}, {Doelman, D. S.}, {Lozi, J.}, {et~al.} 2019, A\&A, 632, A48, \dodoi{10.1051/0004-6361/201936062}

\bibitem[{Corrigan {et~al.}(2018)Corrigan, Morris, Harris, \& Anagnos}]{Corrigan:18}
Corrigan, M.~K., Morris, T.~J., Harris, R.~J., \& Anagnos, T. 2018, in Adaptive Optics Systems VI, ed. L.~M. Close, L.~Schreiber, \& D.~Schmidt, Vol. 10703, International Society for Optics and Photonics (SPIE), 107035H, \dodoi{10.1117/12.2311336}

\bibitem[{Engler {et~al.}(2022)Engler, Louarn, Verinaud, Weddell, \& Clare}]{Engler:22}
Engler, B., Louarn, M., Verinaud, C., Weddell, S., \& Clare, R. 2022, Journal of Astronomical Telescopes, Instruments, and Systems, 8, \dodoi{10.1117/1.JATIS.8.2.021502}

\bibitem[{Jovanovic {et~al.}(2016)Jovanovic, Schwab, Cvetojevic, Guyon, \& Martinache}]{Jovanovic:16}
Jovanovic, N., Schwab, C., Cvetojevic, N., Guyon, O., \& Martinache, F. 2016, Publications of the Astronomical Society of the Pacific, 128, 121001, \dodoi{10.1088/1538-3873/128/970/121001}

\bibitem[{Jovanovic {et~al.}(2023)Jovanovic, Gatkine, Anugu, Amezcua-Correa, Basu~Thakur, Beichman, Bender, Berger, Bigioli, Bland-Hawthorn, Bourdarot, Bradford, Broeke, Bryant, Bundy, Cheriton, Cvetojevic, Diab, Diddams, Dinkelaker, Duis, Eikenberry, Ellis, Endo, Figer, Fitzgerald, Gris-Sanchez, Gross, Grossard, Guyon, Haffert, Halverson, Harris, He, Herr, Hottinger, Huby, Ireland, Jenson-Clem, Jewell, Jocou, Kraus, Labadie, Lacour, Laugier, Ławniczuk, Lin, Leifer, Leon-Saval, Martin, Martinache, Martinod, Mazin, Minardi, Monnier, Moreira, Mourard, Nayak, Norris, Obrzud, Perraut, Reynaud, Sallum, Schiminovich, Schwab, Serbayn, Soliman, Stoll, Tang, Tuthill, Vahala, Vasisht, Veilleux, Walter, Wollack, Xin, Yang, Yerolatsitis, Zhang, \& Zou}]{roadmap}
Jovanovic, N., Gatkine, P., Anugu, N., {et~al.} 2023, Journal of Physics: Photonics.
\newblock \url{http://iopscience.iop.org/article/10.1088/2515-7647/ace869}

\bibitem[{Korkiakoski {et~al.}(2014)Korkiakoski, Keller, Doelman, Kenworthy, Otten, \& Verhaegen}]{FF}
Korkiakoski, V., Keller, C.~U., Doelman, N., {et~al.} 2014, Appl. Opt., 53, 4565, \dodoi{10.1364/AO.53.004565}

\bibitem[{Leon-Saval {et~al.}(2005)Leon-Saval, Birks, Bland-Hawthorn, \& Englund}]{Leon-Saval:05}
Leon-Saval, S.~G., Birks, T.~A., Bland-Hawthorn, J., \& Englund, M. 2005, Opt. Lett., 30, 2545, \dodoi{10.1364/OL.30.002545}

\bibitem[{Lin {et~al.}(2022)Lin, Fitzgerald, Xin, Guyon, Leon-Saval, Norris, \& Jovanovic}]{Lin:22}
Lin, J., Fitzgerald, M.~P., Xin, Y., {et~al.} 2022, J. Opt. Soc. Am. B, 39, 2643, \dodoi{10.1364/JOSAB.466227}

\bibitem[{Lin {et~al.}(2023)Lin, Fitzgerald, Xin, Kim, Guyon, Leon-Saval, Norris, \& Jovanovic}]{Lin:23}
---. 2023, J. Opt. Soc. Am. B

\bibitem[{Lin {et~al.}(2021)Lin, Jovanovic, \& Fitzgerald}]{Lin:21}
Lin, J., Jovanovic, N., \& Fitzgerald, M.~P. 2021, J. Opt. Soc. Am. B, 38, A51, \dodoi{10.1364/JOSAB.423664}

\bibitem[{Lozi {et~al.}(2020)Lozi, Guyon, Vievard, Sahoo, Deo, Jovanovic, Norris, Martinod, Mazin, Walter, Fruitwala, Steiger, Davis, Tuthill, Kudo, Kawahara, Kotani, Ireland, Anagnos, Schwab, Cvetojevic, Huby, Lacour, Barjot, Groff, Chilcote, Kasdin, Martinache, Laugier, N'Diaye, Knight, Males, Bos, Snik, Doelman, Miller, Bendek, Belikov, Pluzhnik, Currie, Kuzuhara, Uyama, Nishikawa, Murakami, Hashimoto, Minowa, Clergeon, Ono, Takato, Tamura, Takami, \& Hayashi}]{Lozi:20}
Lozi, J., Guyon, O., Vievard, S., {et~al.} 2020, in Adaptive Optics Systems VII, ed. L.~Schreiber, D.~Schmidt, \& E.~Vernet, Vol. 11448, International Society for Optics and Photonics (SPIE), 114480N, \dodoi{10.1117/12.2562832}

\bibitem[{{Martinache}(2013)}]{APFWFS}
{Martinache}, F. 2013, pasp, 125, 422, \dodoi{10.1086/670670}

\bibitem[{{Martinez, P.} {et~al.}(2013){Martinez, P.}, {Kasper, M.}, {Costille, A.}, {Sauvage, J. F.}, {Dohlen, K.}, {Puget, P.}, \& {Beuzit, J. L.}}]{Martinez:13:NCPA2}
{Martinez, P.}, {Kasper, M.}, {Costille, A.}, {et~al.} 2013, A\&A, 554, A41, \dodoi{10.1051/0004-6361/201220820}

\bibitem[{{Martinez, P.} {et~al.}(2012){Martinez, P.}, {Loose, C.}, {Aller Carpentier, E.}, \& {Kasper, M.}}]{Martinez:12:NCPA1}
{Martinez, P.}, {Loose, C.}, {Aller Carpentier, E.}, \& {Kasper, M.} 2012, A\&A, 541, A136, \dodoi{10.1051/0004-6361/201118459}

\bibitem[{{National Academies of Sciences, Engineering, and Medicine}(2021)}]{decadal}
{National Academies of Sciences, Engineering, and Medicine}. 2021, Pathways to Discovery in Astronomy and Astrophysics for the 2020s (Washington, DC: The National Academies Press), \dodoi{10.17226/26141}

\bibitem[{{N'Diaye} {et~al.}(2014){N'Diaye}, {Dohlen}, {Caillat}, {Costille}, {Fusco}, {Jolivet}, {Madec}, {Mugnier}, {Paul}, {Sauvage}, {Soummer}, {Vigan}, \& {Wallace}}]{ZELDA}
{N'Diaye}, M., {Dohlen}, K., {Caillat}, A., {et~al.} 2014, in Society of Photo-Optical Instrumentation Engineers (SPIE) Conference Series, Vol. 9148, Adaptive Optics Systems IV, ed. E.~{Marchetti}, L.~M. {Close}, \& J.-P. {Vran}, 91485H, \dodoi{10.1117/12.2056818}

\bibitem[{{N\'{}Diaye, M.} {et~al.}(2018){N\'{}Diaye, M.}, {Martinache, F.}, {Jovanovic, N.}, {Lozi, J.}, {Guyon, O.}, {Norris, B.}, {Ceau, A.}, \& {Mary, D.}}]{NDiaye:18}
{N\'{}Diaye, M.}, {Martinache, F.}, {Jovanovic, N.}, {et~al.} 2018, A\&A, 610, A18, \dodoi{10.1051/0004-6361/201731985}

\bibitem[{Norris {et~al.}(2020)Norris, Wei, Betters, Wong, \& Leon-Saval}]{Norris:20}
Norris, B. R.~M., Wei, J., Betters, C.~H., Wong, A., \& Leon-Saval, S.~G. 2020, Nature Communications, 11, 5335, \dodoi{10.1038/s41467-020-19117-w}

\bibitem[{Sauvage {et~al.}(2007)Sauvage, Fusco, Rousset, \& Petit}]{Sauvage:07}
Sauvage, J.-F., Fusco, T., Rousset, G., \& Petit, C. 2007, J. Opt. Soc. Am. A, 24, 2334, \dodoi{10.1364/JOSAA.24.002334}

\bibitem[{Sauvage {et~al.}(2012)Sauvage, Mugnier, Paul, \& Villecroze}]{COFFEE}
Sauvage, J.-F., Mugnier, L., Paul, B., \& Villecroze, R. 2012, Opt. Lett., 37, 4808, \dodoi{10.1364/OL.37.004808}

\bibitem[{{Skaf, Nour} {et~al.}(2022){Skaf, Nour}, {Guyon, Olivier}, {Gendron, \'Eric}, {Ahn, Kyohoon}, {Bertrou-Cantou, Arielle}, {Boccaletti, Anthony}, {Cranney, Jesse}, {Currie, Thayne}, {Deo, Vincent}, {Edwards, Billy}, {Ferreira, Florian}, {Gratadour, Damien}, {Lozi, Julien}, {Norris, Barnaby}, {Sevin, Arnaud}, {Vidal, Fabrice}, \& {Vievard, S\'ebastien}}]{DRWHO}
{Skaf, Nour}, {Guyon, Olivier}, {Gendron, \'Eric}, {et~al.} 2022, A\&A, 659, A170, \dodoi{10.1051/0004-6361/202141514}

\bibitem[{{Traub} \& {Oppenheimer}(2010)}]{Traub:10}
{Traub}, W.~A., \& {Oppenheimer}, B.~R. 2010, in Exoplanets, ed. S.~{Seager}, 111--156

\bibitem[{{Wilby} {et~al.}(2017){Wilby}, {Keller}, {Snik}, {Korkiakoski}, \& {Pietrow}}]{cMWFS}
{Wilby}, M.~J., {Keller}, C.~U., {Snik}, F., {Korkiakoski}, V., \& {Pietrow}, A.~G.~M. 2017, aap, 597, A112, \dodoi{10.1051/0004-6361/201628628}

\end{thebibliography}
\bibliographystyle{aasjournal}



\end{document}